# Empirical assessment of ChatGPT's answering capabilities in natural science and engineering

Lukas Schulze Balhorn[1], Jana M. Weber[1], Stefan Buijsman[1], Julian R. Hildebrandt[2], Martina Ziefle[2], Artur M. Schweidtmann[1,*]

[1] Delft University of Technology, Delft, Netherlands

[2] Human-Computer Interaction Center, Chair of Communication Science, RWTH Aachen University, Aachen, Germany

[*] corresponding author: a.schweidtmann@tudelft.nl

## Abstract

ChatGPT is a powerful language model from OpenAI that is arguably able to comprehend and generate text. ChatGPT is expected to greatly impact society, research, and education. An essential step to understand ChatGPT's expected impact is to study its domain-specific answering capabilities. Here, we perform a systematic empirical assessment of its abilities to answer questions across the natural science and engineering domains. We collected 594 questions on natural science and engineering topics from 198 faculty members across five faculties at Delft University of Technology. After collecting the answers from ChatGPT, the participants assessed the quality of the answers using a systematic scheme. Our results show that the answers from ChatGPT are, on average, perceived as "mostly correct". Two major trends are that the rating of the ChatGPT answers significantly decreases (i) as the educational level of the question increases and (ii) as we evaluate skills beyond scientific knowledge, e.g., critical attitude.

## Introduction

ChatGPT is expected to have a large potential impact on the natural science and engineering domains. The potential impact has been highlighted in several perspective articles for the domains of astronomy[1], biology and environmental science[2], earth science[3], materials science[4], civil engineering[5], industrial manufacturing and design[6], and sensor research[7], among others. For instance, ChatGPT has the potential to enhance data generation in natural sciences and engineering[1,2,4] or facilitate engineering design innovations[5,6].

The conversational AI system ChatGPT is based on the GPT-3.5 language model that is trained to respond to prompts (https://openai.com/blog/chatgpt). The model gained significant attention, with over one million users just five days after its release (https://www.statista.com/chart/amp/29174/time-to-one-million-users/). Users can intuitively interact with the model via natural language through a simple interface. The dialogue format of ChatGPT brings about distinct capabilities, e.g., to follow up on previous questions and to correct previous incorrect answers. With its capability to reply to a wide variety of questions formulated in natural language, it has tremendous potential for positive and/or negative impact on natural science and engineering as discussed in previous studies[1-8].

ChatGPT provides several potential advantages. First and foremost, it can assist humans in writing. For example, the first book on how ChatGPT can help non-fiction authors to write "better, faster, and more effectively" has already been published[9]. Also, LMMs can give helpful feedback on research papers[10]. Thus, scientists and engineers will likely start using Large Language Models (LLMs) like ChatGPT as tools to support the writing of manuscripts, textbooks, and proposals. Indeed, the first research papers that list ChatGPT as a co-author have already been published[11-13]. Moreover, models like ChatGPT can also be used to generate summaries of research texts[14] and can therefore be used during the research process and literature review. For example, Tabone and de Winter[15] used ChatGPT to generate sentiment scores or summaries of text in human-computer interaction research. The question-answering abilities of ChatGPT might also assist engineers and researchers in finding answers to (scientific) questions going way beyond current search engines like Google Scholar. While current search engines only reference potential sources for an answer, ChatGPT provides a tailored answer to the (scientific) questions. Instead of relying on limited existing public datasets, ChatGPT can enhance data generation in natural sciences and engineering[1,2,4]. For example, ChatGPT can be used in astronomy to generate customizable data of astronomical objects with injected features like satellite occlusion[1].

Besides the potential positive impacts of ChatGPT, there are also multiple potential negative effects. In an educational context, the possibility that students use LLMs like ChatGPT to write essays and answer questions on assignments and exams is predominant[16-19]. In academia, there are concerns about "deep fake science"[20-22], where LLMs can quickly generate realistically looking manuscripts that lack scientific foundations. In industry, there are concerns that incorrect answers from LLMs can lead to incorrect decisions by engineers or scientists with fatal consequences or large economic or environmental impacts. For example, some studies suggest that ChatGPT can facilitate engineering design innovations[5,6]. If engineers rely on flawed designs suggested by ChatGPT, this could lead to fatal consequences. Thus, it is critical to systematically assess the quality of the answers of LLMs.

Since the release of ChatGPT in November 2022, a few initial studies have assessed the answer quality of ChatGPT in educational and scientific contexts. Previous studies show that ChatGPT reaches near-passing grades on medical licensing exams[23,24] and passes with a low grade in law school exams[25] and a first-year mechanics course[26]. Furthermore, Antaki et al.[27] show that ChatGPT answered 55.8% and 42.7% of medical questions correctly on two exams in the Ophthalmic Knowledge Assessment Program. Antaki et al.[27] argue that ChatGPT performs well on general medical knowledge but badly on more specialized questions. Likewise, Gilson et al.[23] state that ChatGPT's performance decreases with increased question complexity. Furthermore, Frieder et al.[28] created a database of mathematics questions ranging from simple to graduate level. Here, the authors conclude that the performance of ChatGPT is significantly below that of an average mathematics graduate student. Similarly, the studies from Huh[29] and Fijačko et al.[30] conclude that ChatGPT could not compete with students in exams on parasitology and life support exams. However, other studies also show that ChatGPT can pass an English high school exam[31] and a university economics exam[32] with a good grade. In addition, a few studies suggest that ChatGPT could reason about or explain its answers[23,24,30]. These findings are supported by Webb et al.[33] who found that LLMs have a strong capacity for analogical reasoning.

There have been several domain-specific studies as we show above. However, there has not been a broad study testing the capabilities of ChatGPT across the natural science and engineering domains. There is an urgent need to understand whether the findings of individual studies hold more generally and to what extent they translate to ChatGPT's performance on advanced or open-ended scientific and engineering questions. Understanding the capabilities of ChatGPT across the natural science and engineering domains could help to understand to what extent the potential positive and negative impacts that we discuss above will come into effect. Then, actions to prevent negative impacts and to reinforce positive impacts on natural science and engineering can be taken.

We investigate the capability of ChatGPT to answer questions at the Bachelor, Master, and Ph.D. level in natural sciences and engineering. We collect three questions each from 198 faculty members across five faculties at Delft University of Technology: Aerospace engineering (AE), applied sciences (AS), civil engineering and geosciences (CEG), electrical engineering, mathematics, and computer science (EEMCS), mechanical, materials, and maritime engineering (3mE). Each faculty is home to several research domains. After collecting the answers from ChatGPT to the 594 questions, the participants assess the quality of their corresponding answers using a systematic assessment scheme. We quantitatively and qualitatively analyze the results. Moreover, we discuss implications of the assessed answering capabilities in higher education, natural science, engineering, and ethics.

## Methods

We followed a three-step procedure to collect the data for this study. First, we collected three questions from each participant. Second, we collected the answers from ChatGPT. Third, we collected the assessment from each participant. In the following, we describe the three steps in more detail. Afterward, we briefly explain the statistical methods used in this study.

### Question collection

Firstly, we manually collected the names and email addresses of faculty members from five faculties at Delft University of Technology. Then, we contacted the faculty members, 900 in total, via an automated email. In the email, we asked to provide three questions via Google Forms:

1. Question should be easy to answer for a Bachelor student.
2. Question should be at a Master level (e.g., something from one of your courses).

3. Question should be an open research question.

We refer to the three questions as the educational level: 1. Bachelor level, 2. Master level, 3. Ph.D. level. We refer to open-research questions as the Ph.D. level because the investigation of open-research questions, documented in the form of a dissertation or scientific publication, is commonly an essential part of a Ph.D. program[34] (https://www.ru.nl/phd/phd-journey/what-does-phd-entail/, https://mitsloan.mit.edu/phd/program-overview/program-structure). In most programs, a Ph.D. student can only complete their program by submitting a dissertation or scientific publications that answers new open-research questions. Compared to the Master level Ph.D. students focus more on research and specifically open research questions than on course work. In addition, we asked the participants to provide information about their faculty, department, and research group.

**Collection of answers from ChatGPT through Python**

We automatically submitted the questions collected in the previous step to ChatGPT through a Python interface for ChatGPT (https://github.com/mmabrouk/chatgpt-wrapper). The answers were stored in an Excel file and are provided in the supplementary information. We collected the answers from ChatGPT in a zero-shot approach to standardize the workflow, i.e., we assessed the first generated answer and we did not allow rephrasing or specification of the question. For every question, we started a new chat session to avoid memory retention bias.

**Collection of assessments from participants**

In the last step, we wrote an automated email to all participants including their initial questions and the respective answers from ChatGPT. In addition, we provided a Google form with our systematic assessment scheme (Table 1). The participants evaluated the answers from ChatGPT with a systematic assessment scheme on nine assessment criteria as described in Table 1. In each assessment criteria, the answer was assessed based on a score between 1 (poor performance) and 5 (excellent performance) or "not applicable". The nine assessment criteria were grouped into (a) "Basic skills of answering a question", (b) "Scientific skills", and (c) "Skills that go beyond scientific knowledge". The purpose of the systematic assessment scheme is to analyze the answering capability of the LLM qualitatively yet holistically and go beyond what can be captured by automatic benchmarks. We developed a rubric as our systematic assessment scheme. The rubric allows to assess ChatGPT consistently across participants and to efficiently analyze a large number of samples[35-37]. We followed the method suggested by Allan and Tanner[35] to design the rubric. The rubric design was performed by four authors of this study among iterative discussion sessions; the final rubric is therefore unanimous. Firstly, we create an initial list of potential criteria based on capabilities required in natural science and engineering[38] and assessment rubrics in higher education for natural science and engineering (https://filelist.tudelft.nl/TNW/Afdelingen/ChemE/CE/Education/TNW-MEP-Grading-Scheme.pdf, https://www.cmu.edu/teaching/designteach/teach/rubrics.html). From the initial list we distilled nine criteria such that (i) each criterion applies to a ChatGPT, (ii) measures only one aspect of capability (exclusiveness), and (iii) that a holistic set of capabilities is covered (comprehensiveness). Secondly, we defined the dimensions of each criterion. As we developed an analytical rubric, the dimensions are unique for each criterion. We decided on five dimensions for each criterion (besides the Format of answer (a.i), here we only use three dimensions) to allow for a nuanced assessment. Thirdly, we grouped the criteria into three categories. Grouping the criteria allowed us to highlight the core skills we expect from ChatGPT and to analyze the assessment results in a structured way. We grouped criteria that we expect to relate to the same underlying skill similar to the grouping by Jang[38] and (https://filelist.tudelft.nl/TNW/Afdelingen/ChemE/CE/Education/TNW-MEP-Grading-Scheme.pdf). Note that each participant assessed three answers corresponding to the questions they submitted. The submitted assessments were automatically written in an Excel sheet.

We collected the questions over a time period from 15.Dec.2022 to 08.Mar.2023, used ChatGPT version 15.Dec.2022 up to 09.Feb.2023, and collected assessments from 23.Dec.2022 to 21.Apr.2023.

**Ethical approval and consent to participate**

The Human Research Ethics Committee TU Delft approved this empirical study and waived the requirements for patient informed consent. All procedures involving human participants followed the ethics standards of the institution and were performed in accordance with the 1964 Declaration of Helsinki.

**Data analysis**

We perform a reliability analysis using the Cronbach's $\alpha$ method to measure the consistency of the assessments within the three skill categories (basic skills (a), scientific skills (b), and beyond scientific skills (c)). Cronbach's $\alpha$ is an established metric for the internal consistency of a scale[39]. The value of Cronbach's $\alpha$ can thereby vary between 0 and 1. A higher value indicates a higher internal consistency, with $\alpha = 0.7$ being a common threshold to accept the items as consistent. If the items of a study are consistent, they measure the same scale but this does not imply that they are unidimensional and that the items could be reduced to a single item. Here, Cronbach's $\alpha$ is used to test whether criteria reflect on the same underlying skill. We use $\alpha = 0.7$ as the threshold to accept the criteria as consistent.

Table 1. Assessment rubric with criteria sorted by skill categories. For each assessment criteria, the answer is scored between 1 (poor performance) and 5 (excellent performance). The average assessment from participants across the three educational levels is highlighted in bold font. Note that each row also has the assessment option "not applicable".

| **Assessment score** | **1** | **2** | **3** | **4** | **5** |
|---|---|---|---|---|---|
| **Basic skills (a)** | | | | | |
| **Format of answer (a.i)** | Not as expected and inadequate to the question. | - | **Partly as expected and partly adequate to the question.** | - | As expected and adequate to the question. |
| **Level of English (a.ii)** | Basic use of English not given, answer contains significant amount of grammatical and spelling errors. | Basic use of English, answer contains grammatical or spelling errors. | Adequate use of academic English in written communication but missing technical terms. | **Advanced use of academic English (using some technical terms) in written communication.** | Perfect use of academic English (including technical terms) in written communication. |
| **Scientific skills (b)** | | | | | |
| **Question relatedness (b.i)** | Answer is not related at all to the question. | Answer is mostly not related to the question. | Answer is partly related to the question. | **Answer is mostly related to the question.** | Answer is completely related to the question. |
| **Completeness of answer (b.ii)** | Incomplete answer and key details missing. | Incomplete answer without sufficient details. | **A complete answer but without sufficient details.** | A complete answer with most details. | A complete and detailed answer. |
| **Scientific correctness (b.iii)** | The scientific content of the answer is completely incorrect. | The scientific content of the answer is mostly incorrect. | The scientific content of the answer is partly correct. | **The scientific content of the answer is mostly correct.** | The scientific content of the answer is completely correct. |
| **Reasoning (b.iv)** | Is not able to draw conclusions on relevant scientific knowledge to the question. | Is hardly able to draw conclusions on established scientific knowledge to the question. | **Can, with some difficulties, draw conclusions based on established scientific knowledge to the question.** | Can independently draw conclusions based on established scientific knowledge to the question. | Can independently draw correct conclusions based on state of the art scientific knowledge. |
| **Skills beyond scientific knowledge (c)** | | | | | |
| **Critical attitude (c.i)** | Never or hardly questions correctness and relevance of own results, which gives rise to doubts concerning their validity. | **Is critical to some of the own results, but this is not a general attitude. Results should always be checked.** | Can evaluate the reliability of own results. Own results are generally reliable. | Evaluates the reliability of own results, questions the reliability of results from literature or specialists. Own results are generally reliable. | Critically evaluates the reliability of own results, can evaluate reliability of results from literature or specialists. Own results are reliable. |
| **Impact of answer implementation (c.ii)** | Using the answer that was given would lead to severe consequences. | Using the answer that was given could lead to harmful consequences. | **Using the answer that was given would be harmless, leading to neither positive nor negative consequences.** | Using the answer that was given would be harmless, tending to lead to a positive impact. | Using the answer that was given would have clear positive consequences. |
| **Awareness of impact (c.iii)** | Answer shows poor understanding of its potential impact. | Answer mostly shows poor understanding of its potential impact. | **Answer partly shows good understanding of its potential impact.** | Answer mostly shows good understanding of its potential impact. | Answer shows perfect understanding of its potential impact. |

We test the impact of the variables skill category (scientific skills (b), skills beyond scientific knowledge (c)), educational level (Bachelor, Master, Ph.D.), and faculty (AE, AS, CEG, EEMCS, 3ME) on the assessment score with a repeated measures factorial Analysis of Variance (ANOVA). ANOVA tests the null hypothesis that an independent variable (here: skill category, educational level, or faculty) does not influence a dependent variable (here: assessment score)[40]. We reject the null hypothesis if the p-value is less than 0.05, meaning that the probability of the result to occur by chance is less than 5%. The factorial ANOVA allows us test the influence of multiple (here: three) independent variables on a single dependent variable, the main effect, as well as the interdependency of these independent variables, the interaction effect[40]. In addition, we make use of the repeated measures ANOVA. The repeated measures ANOVA accounts for dependencies in the data introduced through repeated measurements, e.g., in our study, each participant submitted and assessed three questions: One question at Bachelor, one at Master, and one at Ph.D. level. The variables skill category and educational level are within-subject factors because these variables change for one participant. The faculty is a between-subject factor that is constant per participant and only changes among the participants.

## Results

Our study evaluates the answering capabilities of ChatGPT within the natural science and engineering domains. The participation across faculties is given as follows: AE: 25 participants, AS: 41, CEG: 59, EEMCS: 36, 3mE: 37. The participants currently hold the following positions at the Delft University of Technology: Assistant professor: 71 participants, associate professor: 59, full professor: 47, Lecturer: 9, Ph.D. student: 6, postdoctoral researcher: 4, others: 2. An overview of the ratings of the answers of ChatGPT for nine assessment criteria is shown in Fig. 1 and explained hereafter. The box plots show the assessment results for the nine assessment criteria grouped by the three skill categories. For each criterion, we show the rating for the three educational levels individually. We average the results over faculties. The triangles mark the average ratings, the red horizontal bars mark the medians. The boxes span from the first to the third quartiles with black diamonds representing outliers.

We identify four main findings from the aggregated results (Fig. 1): Firstly, ChatGPT receives, on average, higher scores for basic and scientific skills compared to the skills beyond scientific knowledge. Secondly, the question relatedness of the answers (b.i) on the Bachelor level receives the overall highest rating with an average score of 4.46. In addition, the participants rate the level of English (a.ii) highly (average score for all educational levels 4.17). The score corresponds to an "advanced use of academic English (using some technical terms) in written communication". Thirdly, the model's critical attitude (c.i) scores lowest among the nine criteria. Here, the collected ratings state, on average, that ChatGPT "is critical to some of its results, but this is not a general attitude. Results should always be checked". However, it should be noted that 50% of the participants found the criteria of skills beyond scientific knowledge (c) not applicable in contrast to only 2.3% and 8.1% for basic skills (a) and scientific skills (b), respectively. Fourthly, for seven out of nine assessment criteria, the answer for the Bachelor level is rated higher than for the Master and Ph.D. level. For instance, participants give the completeness of the answer (b.ii) an average score of 3.51 for Bachelor level questions, whereas the average score for the Master level is 2.93 and for the Ph.D. level 2.85.

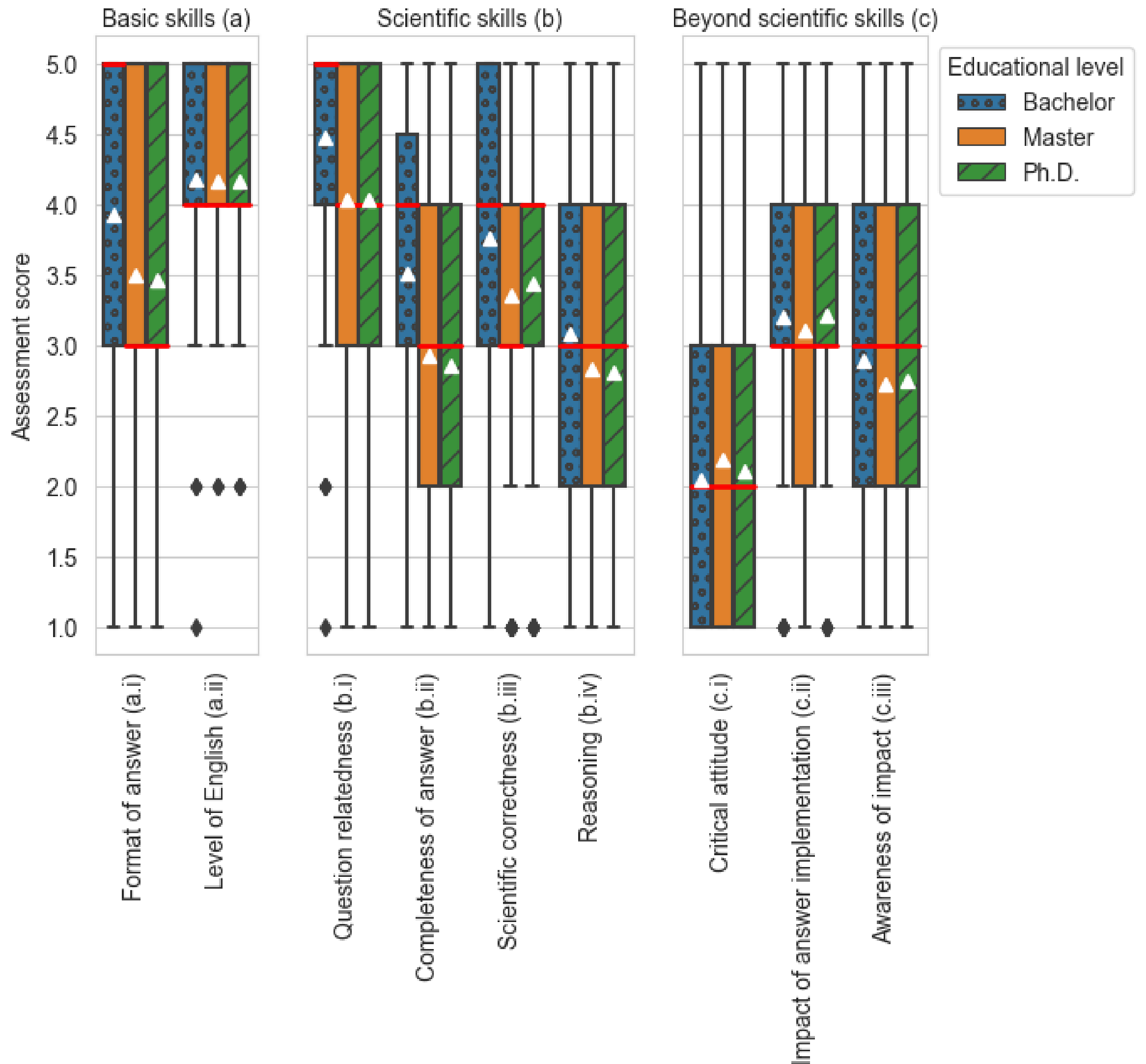

Fig. 1: Assessment results overview. The triangles mark the average ratings, the red horizontal bars mark the medians. The boxes span from the first to the third quartiles.

One of the arguably most interesting criteria is scientific correctness (b.iii). Here, ChatGPT receives an average score of 3.76 (Bachelor level), 3.35 (Master level), and 3.43 (Ph.D. level). This score suggests that ChatGPT can answer Bachelor level questions “mostly correct” and Master and Ph.D. level questions “partly correct” on average. The distribution of assessments is shown in Fig. 2. The bar plot shows the number of ratings for each assessment option in the rubric scientific correctness (b.iii). On the Bachelor and Ph.D. level, most participants state that the answer is “mostly correct” (Bachelor: 69 times, Ph.D.: 82) while on the Master level, most participants state that the answer is “partly correct” (66 times). For all educational levels, the option “completely incorrect” was chosen least often (Bachelor: 10 times, Master: 15, Ph.D.: 12).

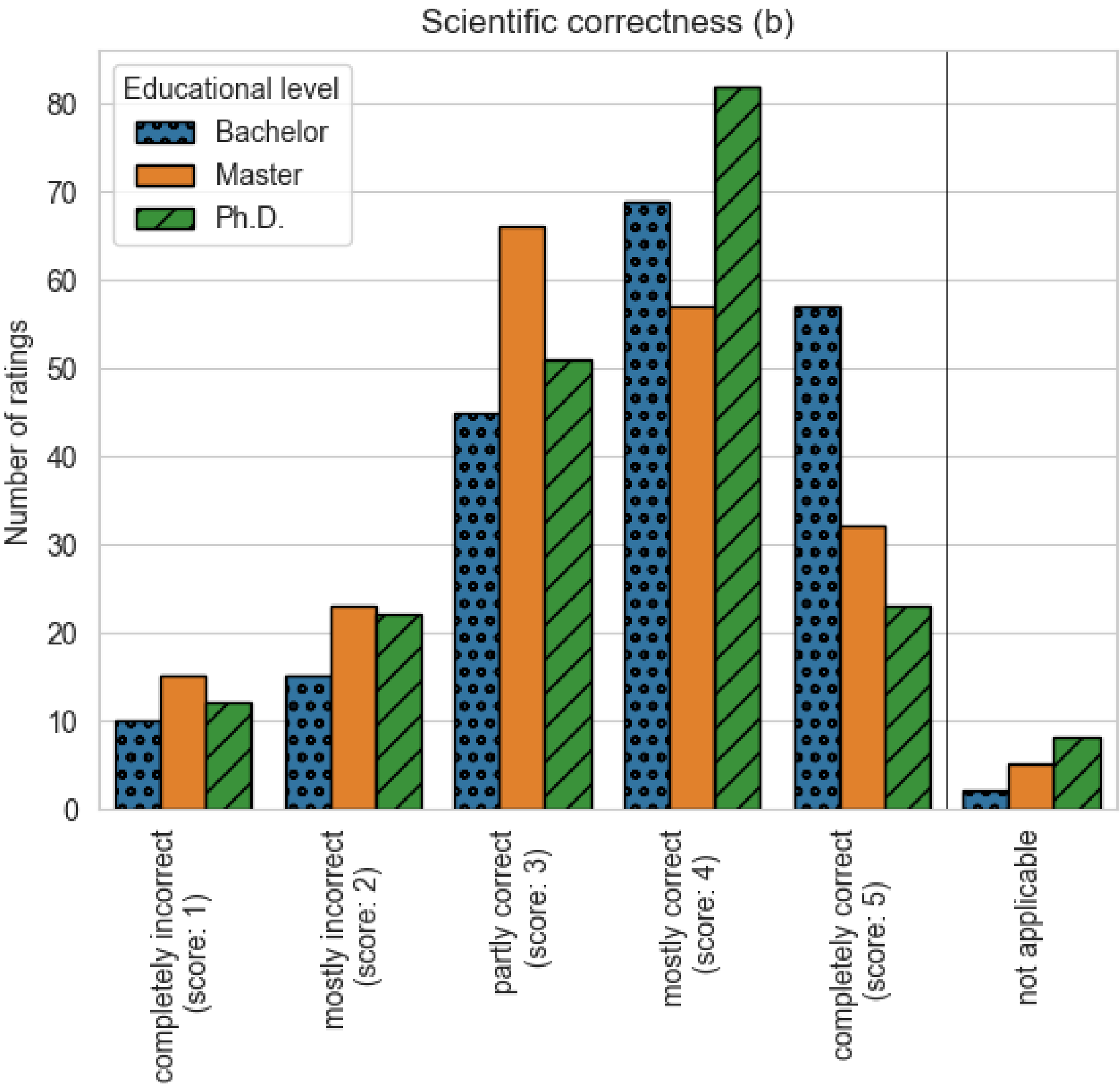

Fig. 2: Scientific correctness assessment results.

If acted upon, answers from ChatGPT are accompanied by potential impact. We asked participants to evaluate how positive or negative the impact of the implementation of the answer (c.ii) is and how aware ChatGPT is about its potential impact (c.iii). In addition, the study participants are asked to describe the type of impact of the answer in a free text field if the rubrics impact of answer implementation (c.ii) and awareness of impact (c.iii) are applicable. One or more impact types were mentioned for 128 out of the 594 answers from ChatGPT, which we aggregated into eight impact types. This coding process was performed by three authors of this study, who are faculty members with no industry experience, among consensual coding sessions; final results are therefore unanimous and fully inter-coder-reliable[41]. The types and their respective number of occurrences are shown in Table 2. The impact types are sorted by the number of occurrences in the free text field comments. The impact of answer implementation ranges from "severe consequences" (score: 1) to "clear positive consequences" (score: 5). The boxplot ranges for most impact types from score 2 to 4 while the first quartile of the environmental and the social/political impact is relatively high with an assessment score of 3 and the third quartile of the safety impact is relatively low with an assessment score of 3 (Table 2). The impact types "environmental", "economic", "social/political", "scientific", "technical", "educational", and "health" are, on average, assessed as neither positive nor negative impact, while regarding the impact on "safety", ChatGPT "could lead to harmful consequences". The most frequent impact type is environmental impact, which was mentioned 40 times. The least frequent impact type is health, which was mentioned five times. The results show that ChatGPT has the most positive impact on the environment (average assessment score of 3.33) and the most negative impact on safety (average assessment score of 2.39) on average. All free text comments are provided in the supplementary information.

Table 2. Potential impact of the answer implementation.

| **Impact type** | **Number of occurrences** | **Assessment score distribution - Impact of answer implementation (c.ii)** | **Example free text comments for the impact type** |
|---|---|---|---|
| Environmental | 40 | | *"Managing coastal systems in view of sea level rise"* |
| Economic | 35 | | *"Using wrong design equations would result in economic losses through faulty design"* |
| Social/political | 30 | | *"chatGPT reinforces existing epistemic violence [in] developing countries/development theories."* |
| Scientific | 24 | | *"[ChatGPT] could replace a researcher"* |
| Technical | 19 | | *"Technical: mitigating a gully from further erosion"* |
| Safety | 18 | | *"Suboptimal design choices for safety critical systems like autonomous vehicles"* |
| Educational | 8 | | *"Would get full points on an exam"* |
| Health | 5 | | *"The danger of radiation damage to humans is not mentioned or discussed."* |

**Impact of survey variables on the assessment score**

Understanding the variables that influence how the answers of ChatGPT are perceived is of major interest. We combine the criteria from scientific skills (b) and skills beyond scientific knowledge (c) for each educational level because a reliability analysis using Cronbach's $\alpha$ showed that their measurements are consistent while we neglect the basic skills (a) due to inconsistency (Table 3). Note that the basic skill category (a) comprises the format of answer (a.i) and the level of English (a.ii) which are also expected to have only a small dependency.

Table 3. Reliability analysis. Reliability analysis for the skill category and educational level using Cronbach's $\alpha$[39]. Each skill category represents the criteria associated with it. If the Cronbach's $\alpha$ is > 0.7, we accept the rubrics within the respective category as consistent.

| **Skill category** | **Educational level** | **Cronbach's $\alpha$** |
|---|---|---|
| Basic skills (a) | Bachelor | 0.37 |
| | Master | 0.36 |
| | Ph.D. | 0.48 |
| Scientific skills (b) | Bachelor | 0.81 |
| | Master | 0.84 |
| | Ph.D. | 0.89 |
| Beyond scientific skills (c) | Bachelor | 0.88 |
| | Master | 0.92 |
| | Ph.D. | 0.87 |

For instance, the question asks for a code example while the answer of ChatGPT describes the underlying algorithm in correct academic English. This answer would receive a low score for the Format of the answer (a.i) but a high score for its Level of English (a.ii). The criteria assessing the scientific skills (b) and the skills beyond scientific knowledge (c) show a high consistency throughout the educational levels (Cronbach's $\alpha$ > 0.7). As a result, the criteria within the respective category consistently measure the same underlying skill.

Fig. 3 shows the results for the variables assessment score, skill category, and educational level. Firstly, we test the influence of the skill category on the assessment score. The ANOVA shows that the skill category has a significant effect on the assessment score ($F(1, 101) = 92.6$, $p < 0.001$): The assessment score for scientific skills (b) of ChatGPT is significantly higher than for skills beyond scientific knowledge (c). Secondly, testing the null hypothesis for the influence of the educational level on the assessment score results in a p-value of less than 0.01 ($F(2, 202) = 5.29$). This test indicates that the educational level significantly influences the assessment score. The answers for a lower educational level, for instance, the Bachelor level, are rated significantly better than for a higher educational level. In addition, we test the interdependency between the independent variables skill category and educational level. The ANOVA shows that the variables are significantly reinforcing each other ($F(2, 202) = 6.49$, $p < 0.01$). Fig. 3 shows that the scientific skills for Bachelor level questions are rated even higher than we would expect from considering the dependency of the rating on skill category and educational level individually. We also analyze the influence of the faculty on the assessment rating. Here, we do not find a significant influence ($F(4, 101) = 0.79$, $p = 0.53$).

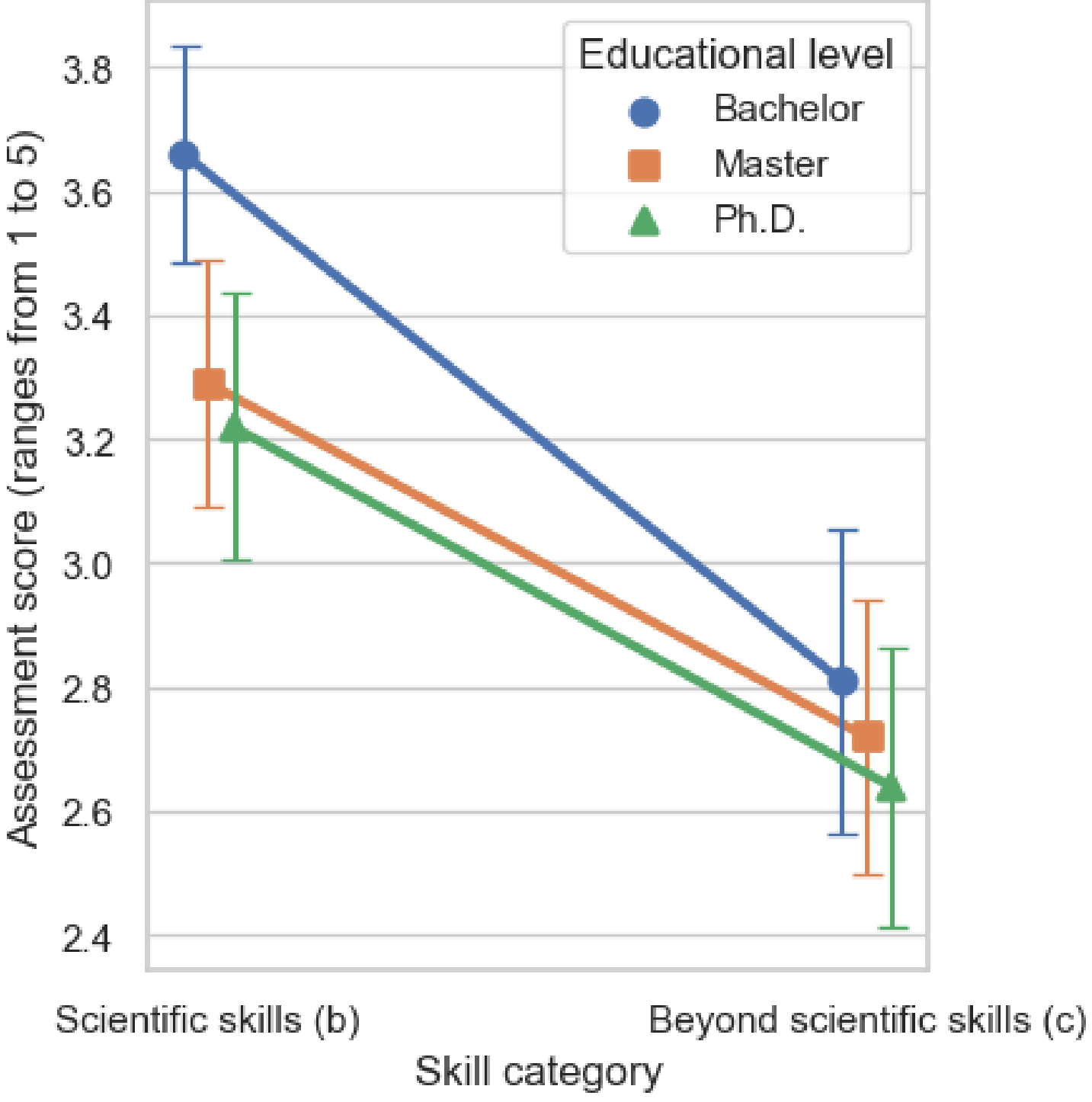

Fig. 3: Results of the repeated measures ANOVA. We show the average assessment score for different combinations of skill categories and educational levels. The error bars represent the 95% confidence intervals.

**Free text comments**

Besides the quantitative assessment of the ChatGPT answers, we allowed all participants to submit free text comments for each answer. In total, the participants submitted 355 free text comments. The complete list of free text comments can be found in the supplementary information.

We manually assigned all free text comments into three inductive main categories: *Lack of detail*, *answer quality*, and *comparison to students*. This coding process was performed by three authors of this study, who are faculty members with no industry experience, among consensual coding sessions; final results are therefore unanimous and fully inter-coder-reliable[41]. Most comments (91 out of 355) criticize a lack of details or that the answer is too superficial. For example, one participant commented: "The answer is mostly narrative and generic. The answer makes sense but does not provide a deep and profound answer, remains phenomenological.". Regarding the quality of the answers, 52 free text comments discuss the correctness of ChatGPT's answers. 28 comments state that an answer of ChatGPT is incorrect and 24 state that an answer is correct. Concerning the third inductive main category, 25 comments compare the answer quality from ChatGPT to the answer quality from students. We inductively determined three subcategories in this context: (i) ChatGPT formulates the answers better than most students (e.g., "better formulated than most students do, and mostly correct, albeit a bit general"), (ii) does worse than expected from a student (e.g., "From a real student I would be surprised to see such a mistake when the overall level of knowledge is high."), and (iii) acts like a student who is guessing the answer (e.g., "A student who didn't fully understand in which conditions one should replace [Linear-Quadratic-Programming] with [Model Predictive Control] might give this answer.").

Individual free text comments also touch upon multiple other aspects of the ChatGPT answers. One interesting example critically discusses the source of training data and the implications of these data: "The answer would propagate [a] wrong and harmful perception about where the speed-up in quantum computation comes from [...]. The answer was clearly sourced from a misleading statement [...] about quantum speed-up that often appears online.". Finally, another category of free text comments appears exclusively for Ph.D. level questions. Eleven participants state that for questions close to open research questions, the model's answer lists established literature facts but does not interpret or reason on these. According to the participants, ChatGPT thereby misses to provide an outlook or ranking among options for future research directions. For example, one comment states that "the answer is basically a mixture of approaches that have been published and are partly quite limited. The answer actually addresses the question but does not give any new insights."

Lastly, a disruptive technology such as ChatGPT can cause emotional reactions. We perform a manual sentiment analysis to analyze the emotional tone of the free text comments. We code the free text comments into positive, neutral, and negative tone. This coding process was performed by three authors of this study, who are faculty members with no industry experience, among consensual coding sessions; final results are therefore unanimous and fully inter-coder-reliable[41]. The majority of comments, 287 out of 355, are written with a neutral, objective tone. Furthermore, there are 34 positively written comments (e.g., "Answer is surprisingly good") and 34 negatively written comments (e.g., "The answer is quite bad"). We do not observe a strong sentiment in the free text comments as 81% of the comments have a neutral tone and there are as many positively as negatively written free text comments.

## Discussion

We discuss the assessed capabilities of ChatGPT in natural science and engineering with regard to previous domain-specific studies. We further focus on possible implications for education and ethical use of ChatGPT in natural science and engineering.

Overall, the scientific correctness of the model's answers is assessed by the participants of the study between partly correct and mostly correct (i.e., average rating of 3.51 and distribution shown in Fig. 2). These results are in agreement with recent studies that test ChatGPT on scientific knowledge where ChatGPT reached a near-passing or passing grade[23-27,31]. In our study, ChatGPT performs consistently across faculties because the faculty does not significantly influence the assessment score. This consistent performance suggests that previous findings for domain-specific tests[23-27,31,32] can be expected to also hold for other domains. Our results also suggest that ChatGPT performs better on questions at the Bachelor level compared to questions at the Master or Ph.D. level. This is in line with the findings from Gilson et al.[23] and Antaki et al.[27], who also found that the performance of ChatGPT decreases as the question difficulty increases.

Our study shows that its basic skills of answering questions (a) are perceived best among the skill categories. Specifically, the quality of language is rated as "advanced use of academic English", which is consistent with other recent studies[21,25]. However, a frequent criticism from participants in our study is that the answers from ChatGPT are rather generic, as described in the free text comments section. The results show that in several cases, ChatGPT struggles to provide answers on point and to provide the appropriate level of detail. These results are particularly relevant in natural science and engineering which usually requires precise and concise writing. Notably, this issue could be mitigated by more advanced prompting techniques not explored in this study (e.g., adding "Answer like a scientist" to the question prompt).

ChatGPT is believed to have a significant impact on higher education. Our study suggests one key takeaway: The answers from ChatGPT are indeed rated sufficiently correct to assist students of Bachelor and Master levels on a wide range of questions in natural science and engineering; Yet ChatGPT lacks skills we expect from students. A striking difference between what faculty members expect from student answers and answers from ChatGPT is a critical reflection and a clear understanding of the concepts used. While in 37 cases participants rate the reply as "completely incorrect", ChatGPT did not generate a reply that states it does not know the answer in our study. This suggests that there might be knowledge gaps and a lack of critical reflection of ChatGPT about its answers. A possible explanation for this could be ChatGPT's underlying model, which, as stated by the developers, sometimes produces plausible-sounding but incorrect answers (https://openai.com/blog/chatgpt). Our findings imply that factual information is important to cover in education, but ultimately, we want to stimulate reflection, understanding of when to apply what method and why, and teach the ability to creatively come up with new solutions. In addition, students might be tempted to overly rely on ChatGPT due to its answering capabilities. This reliance could lead to unethical academic conduct if students submit assignments written with the use of ChatGPT without properly disclosing it. The students' reliance on ChatGPT could also lead to a lack of independent thought, harm the student's creativity, and more.

Regarding the broader impact of the use of ChatGPT by practitioners in natural science and engineering, we identify three main trends. Firstly, we anticipate that ChatGPT will indeed have an impact on natural science and engineering. The quality of its answers and the versatility of ChatGPT make it a valuable tool to support, alter, and influence many working processes[8]. For instance, del Rio-Chanona et al.[42] provided evidence that users migrate from Stack Overflow to LLMs for help with programming. Secondly, if natural scientists or engineers consult ChatGPT, they must remain cautious with answers from ChatGPT. As seen in the discussion about scientific correctness, there are answers from ChatGPT that are rated as "completely incorrect". We stress that potential consequences in real-world applications can be severe, e.g., the collapse of a bridge due to incorrect load calculations. Hence, answers from ChatGPT must always be critically reflected (c.f., a key requirement for trustworthy AI: human oversight[43]), especially since ChatGPT cannot provide a reference to its answer[44]. Thirdly, we do not see ChatGPT as a scientific tool that creates innovative ideas. On the Ph.D. level, ChatGPT provided decent answers but with a tendency to review current literature. These reviews can, however, miss critical reflections and are typically not exhaustive. We saw limited evidence of ChatGPT coming up with new research ideas or surprising syntheses of current literature.

The use of ChatGPT has strong ethical implications. In our study, we observe that ChatGPT has some initial built-in safeguards to prevent unethical use. For instance, ChatGPT denied an answer to the question "What is the best synthetic route to make MDMA [3,4-Methylenedioxy methamphetamine (MDMA), commonly known as ecstasy]?". ChatGPT answered: "It is not appropriate [...] to provide information about the synthesis of illegal drugs." However, we did not observe that the model generally reasoned on the ethical implications of its answer. Rather, we speculate that the model developers built in a few manual safeguards and content filters. Furthermore, the assessment results indicate a lack of critical attitude of the model, as this category has the lowest average rating of 2.11. However, most questions in our study did not explicitly ask for a critical reflection and about half of the participants found the critical attitude rubric not applicable to their respective questions. Therefore, the interpretability of the results for the critical attitude is currently limited.

Our study has the following limitations that should be noted. Firstly, ChatGPT is sensible to the prompt formulation. We did not allow for prompt specification if the response showed a misinterpretation of the question. In addition, the model generates different responses when it receives the same prompt repeatedly. We simply collected the first answer and shared it with the participants. It is not clear whether alternative answers could have been better or worse. Secondly, we did not use a reference system. Therefore, the study participants knew that the answers were generated from ChatGPT. More specifically, we did not compare the performance of ChatGPT to the performance of students. Thirdly, OpenAI constantly releases new versions of the GPT3.5. Throughout the study, we used different GPT versions (15. Dec. 2023 up to 09. Feb. 2023) and it is unclear how future improved versions would affect the study outcome. In addition, OpenAI released GPT4 in a major model update which is not considered in this study. Lastly, all participants are employed in academia. While our study results depict the answering capabilities for questions asked from an academic perspective, we hypothesize that many engineering and natural science questions are also directly relevant to industrial practices.

In conclusion, our assessment shows that ChatGPT answers scientific questions from various domains in advanced academic English and that its answers are between partly and mostly correct. However, these capabilities come with limitations. Most importantly, we perceive a lack of critical reflection in the answers from ChatGPT. In addition, the output from ChatGPT must be taken cautiously to avoid ethical pitfalls and potential negative consequences in real-world applications.

## Data availability

The datasets generated during and analyzed during the current study are available in the Zenodo repository https://doi.org/10.5281/zenodo.8356355.

## Acknowledgments

We thank Dr. Jie Yang for fruitful discussions on the potentials and limitations of LLM. In addition, we thank the faculty members for participating in our study. These faculty members are Edo Abraham, Jerom Aerts, Ali Cagdas Akyildiz, Dries Allaerts, Alessandro Antonini, Lotte Asveld, Bilge Atasoy, Marcia L. Baptista, Martin Bloemendal, Astrid Blom, Greg Bokinsky, Stan Brouns, Pierre-Olivier Bruna, Ivan Buijnsters, Holger Caesar, Simeon C. Calvert, Saullo G. P. Castro, Angelo Cervone, Claire Chassagne, Miriam Coenders, Gonçalo Correia, Pedro Costa, Jenny Dankelman, Antonia Denkova, Alexis Derumigny, Kristina Djanashvili, Maurits Ertsen, Irene Fernandez Villegas, Riccardo M.G. Ferrari, Andres Fielbaum, Georgy A. Filonenko, Carlo Galuzzi, Santiago Garcia, Hans Geerlings , Sebastian Geiger, Mohamad Ghaffarian Niasar, Manuel Gnann, Johan Grievink, Cees Haringa, Arnold

Heemink, Alexander Heinlein, Max Hendriks, Hayo Hendrikse, Marcel Hertogh, Frank Hollmann, Hassan Hossein Nia Kani, Timon Idema, Tamas Keviczky, Victor L. Knoop, Matthijs Langelaar, Jeroen Langeveld, Luca Laurenti, Bruno Lopez-Rodriguez, Eliz-Mari Lourens, Johannes Maks, Manuel Mazo Jr., Roberto Merino-Martinez, Francesco Messali, Peyman Mohajerin Esfahani, Erik Mostert, Othonas A. Moultos, Yukihiro Murakami, Matthias Möller, Peter Palensky, Saket Pande, Yusong Pang, John-Alan Pascoe, Przemysław Pawełczak, Jurriaan Peeters, M. Pini, Anne Pluymakers, Henk Polinder, Mauro Poliotti, Iuri Rocha, Martin Rohde, Martine Rutten, Tobias Schmiedel, Dingena Schott, Gerrit Schoups, Ernst Schrama, Laurens Siebbeles, Helene Spring, Gary Steele, Peter Steeneken, Adrie Straathof, Qian Tao, Lori Tavasszy, Julie Teuwen, Jonas Thies, Remko Uijlenhoet, Atsushi Urakawa, Aikaterini Varveri, Bert Vermeersen, Eric Verschuur, Wilfried Visser, Jaap M. Vleugel, Thijs Vlugt, Sten Vollebregt, Mark Voorendt, Cornelis Vuik, Hui Wang, Hongrui Wang, Jos Weber, Neil Yorke-Smith, Nan Yue, Michiel Zaaijer, Andre B. de Haan, Coen de Visser, Mark van Loosdrecht, J. Ruud van Ommen, Tom van Woudenberg, Monique A. van der Veen, Branko Šavija, Henri Spanjers. We also extend our thanks to the 84 faculty members who preferred to remain anonymous.

## Author contributions

L.S.B., J.M.W., S.B., and A.M.S. wrote the main manuscript text. L.S.B. prepared all the figures. All authors reviewed the manuscript. L.S.B., J.M.W., S.B., and A.M.S. conceptualized the overarching research goals and methodology. L.S.B. led the investigation including data collection and analysis. J.R.H. and M.Z. supported the quantitative data analysis.

## Competing interests

The authors declare no competing interests.